\documentclass[a4paper]{article}
\usepackage{INTERSPEECH2021}
\usepackage{multirow}
\usepackage{stfloats}
\usepackage[moderate]{savetrees}
\usepackage{color,hyperref}
\hypersetup{
    colorlinks = true,
    linkcolor=blue,
    filecolor=blue,      
    urlcolor=blue,
    citecolor=cyan,
}

\title{The 2022 Far-field Speaker Verification Challenge: Exploring domain mismatch and semi-supervised learning under the far-field scenario}
\name{Xiaoyi Qin$^{1}$, Ming Li$^{1}$, Hui Bu$^{4}$,Shrikanth Narayanan$^{3}$, Haizhou Li$^{2}$}

\address{
$^{1}$Data Science Research Center, Duke Kunshan University, Kunshan, China \\
$^{2}$Department of Electrical \& Computer Engineering, National University of Singapore, Singapore \\
$^{3}$Signal Analysis and Interpretation Lab, University of Southern California, Los Angeles, USA \\
$^{4}$AI Shell Foundation, Beijing, China}
\email{ming.li369@duke.edu}

\begin{document}

\maketitle

\begin{abstract}
	FFSVC2022 is the second challenge of far-field speaker verification. To further explore the far-field scenario, FFSVC2022 provides the fully-supervised far-field speaker verification and proposes the semi-supervised far-field speaker verification. In contrast to FFSVC2020, FFSVC2022 focus on the single-channel scenario. In addition, a supplementary set for the FFSVC2020 dataset is released this year. The supplementary set consists of more recording devices and has the same data distribution as the FFSVC2022 evaluation set.  
	This paper summarizes the FFSVC 2022, including tasks description, trial designing details, a baseline system and a summary of challenge results. The challenge results indicate substantial progress made in the field but also present that there are still difficulties with the far-field scenario.
	
\end{abstract}

\section{Introduction}

The success of FFSVC2020 \cite{ffsvc20} indicates that more and more researchers are paying attention to the far-field speaker verification task. This year, the challenge is still focusing on the far-field speaker verification scenario and providing a new far-field development and evaluation set collected by real speakers in complex environments with multiple scenarios, e.g., text-dependent, text-independent, cross-channel enroll-test, multiple visiting recording etc. In addition, speech data is not always labeled in the real scenario, especially for far-field data, which is also hard to accurately labeled with a close-talking model with good discriminatory. Therefore, a new focus of this year is cross-language self-supervised/semi-supervised learning, where the speaker labels of in-domain data (FFSVC2020 dataset, in Mandarin) are not allowed. We encourage participants to generate the pseudo-label for the train/dev set without using the speaker label of the FFSVC2020 dataset (in Mandarin) by the close-talking model trained by VoxCeleb1\&2 (mostly in English) to fine-tune the model.

In contrast to FFSVC2020 tasks, this challenge focuses on the single-channel scenario, which means that both the enrollment and test audio are single-channel data. In addition, considering the real application scenario, this year's trial list will be more challenging than FFSVC2020. The trial pairs will consider more hard cases, e.g.,  text-mismatch, cross-domain, cross-channel, cross-time, etc. 

The FFSVC2022 is designed to boost the speaker verification research with special focus on far-field scenario under noisy conditions in real scenarios. The objectives of this challenge are to: 1) benchmark the current speech verification technology under this challenging condition, 2) promote the development of new ideas and technologies in speaker verification, 3) provide a real, hard, and exploring challenge to the community that exhibits the far-field characteristics in real scenes. The new official challenge website has been published in \href{https://ffsvc.github.io}{https://ffsvc.github.io}. 

\section{Tasks Description}

This year we focus on the far-field single-channel scenarios. There are two tasks in this challenge; both tasks are to determine whether two speech samples are from the same speaker:
\begin{itemize}
	\item \textbf{Task 1. Fully supervised far-field speaker verification.}
	\item \textbf{Task 2. Semi-supervised far-field speaker verification.}
\end{itemize}

Task 1 is the fully supervised far-field speaker verification task that using fixed large-scale close-talking databases and FFSVC20 dataset.

Task 2 is the semi-supervised far-field speaker verification task that participants are allowed to use fixed large-scale close-talking databases with speaker labels and the FFSVC20 dataset without speaker labels.

\subsection{Trial cases}

Both tasks adopt the same trial file. The following cases will be considered in the final trials.

\begin{itemize}
	\item Gender. For negative pairs, most negative trial audios are selected from the same gender, and also, a few cross-gender trial pairs are provided.
	\item Cross-domain. Close-talking or far-field speech segment is chosen as test audio, and enrollment data uses the close-talking speech segment.
	\item Cross-channel. Telephone recorded audios are chosen as enrollment data, and tablet/telephone recorded audios are chosen as test data.
	\item Cross-time. Since the recording process lasted one month. Each speaker has three-time visits, and each visit has a 7-15 days time span. Therefore, enrollment and test data are chosen from different visits data.
	\item Text-mismatch. The trial consists of the text-dependent and text-independent trial pairs.
\end{itemize}

The final trials consist of the combination of the mentioned above cases. The participants are expected to explore more novel and robust systems.

\subsection{Training data}

We define the task 1\&2 as fixed training conditions that the participants can only use special training set to build a speaker verification system. The fixed training set consists of the following:
\begin{itemize}
	\item VoxCeleb 1\&2 \cite{vox1,vox2}.
	\item FFSVC2020 dataset (Train and dev set) \cite{ffsvc20,ffsvc20_plan}. 
	\item FFSVC2020 supplementary set (released in this challenge). This dataset is the supplement of FFSVC20 dataset, both datasets consist of the same speakers. 
\end{itemize}  

Since the FFSVC2022 adopts tablet/telephone data as evaluation data, we release a supplementary set of FFSVC2020, which also consists of the tablet/telephone data to reduce the negative effect of the unseen channel. 

For task 1, participants can only use the VoxCeleb 1\&2 dataset and the train/dev/supplementary sets of the FFSVC2020 dataset with speakers labeled to train the model.

For task 2, in contrast to task1, \textbf{participants cannot use the speaker label of the FFSVC2020 dataset}. In this task, we encourage the participants to adopt the self-supervised or semi-supervised methods to solve the problem of cross-domain unlabeled data, e.g., identity pseudo-label using the pre-trained model on the VoxCeleb dataset.

Using any other speech data in training is forbidden, while participants are allowed to use the non-speech data to do data augmentation. The self-supervised pre-trained models, such as Wav2Vec \cite{wav2vec} and WavLM \cite{wavlm}, cannot be used in this challenge.

\subsection{Development set}

This year, we publish new development and evaluation sets, which are selected from the DMASH dataset\footnote{https://www.aishelltech.com/DMASH\_Dataset}. The trial file and wav files with accurate speaker information will be provided for participants as the development set. The development data has the same data distribution as evaluation data. However, the development set is only allowed to tune hyperparameters and test model performance. Any circumstances of training development set are not allowed, e.g., using development set to train PLDA or speaker system.

\subsection{Evaluation set}

As mentioned before, the evaluation set is not completely out-domain data. However, unlike the FFSVC2020 challenge, we introduce new recording devices, and all trial pairs are single-channel speech segments. The evaluation set consists of a large trials file and anonymized audios. 

\section{Evaluation Protocol}

In this challenge, we will use two metric to evaluate the system performance. The primary metric we adopt is the Minimum Detection Cost(mDCF). Equal Error Rate (EER) will be provided to participant as auxiliary metrics.

The mDCF is based on the following detection cost function which is the same setting as used in the NIST 2010 SRE. It is a weighted sum of miss and false alarm error probabilities in the form:

\begin{equation}
C_{det}=C_{miss}\times P_{miss}\times P_{tar} + C_{fa}\times P_{fa}\times (1-P_{tar})
\end{equation}

We assume a prior target probability, P$_{tar}$ of 0.01 and equal costs between misses and false alarms. The model parameters are 1.0 for both C$_{miss}$ and C$_{fa}$. 

\section{Baseline System}

\subsection{Task1: Fully-supervised Learning}

%\begin{figure}
%  \centering
%  \includegraphics[width=\linewidth]{./fine-tune_s}
%  \caption{Adaptation pipeline of the transfer learning strategy. }
%  \label{fig:finetune}
%\end{figure} 

The task1 baseline system adopts a transfer learning training strategy named FT-mix. The FT-mix training strategy uses the large-scale out-of-domain dataset (5994 speakers of VoxCeleb2 development set) to train a pre-train model and fine-tune it using mixed data of both in-domain data (120 speakers of FFSVC2020 supplementary set) and out-of-domain data. The detail implement is describe following,
% The adaptation process of the transfer learning strategy is shown in Fig. \ref{fig:finetune}. 
 \begin{itemize}
	\item Pre-train a deep speaker embedding model using the Vox2Dev dataset with 5994 speakers;
	\item Retain all parameters of the model except for the output speaker classification layer (num = 5994); replace the speaker classifier with respect to the number of speakers in the mixed data (num=5994+120);
	\item Fine-tune and adapt the the new model with the mixed data until it converges. All parameters, including those from the pre-trained model and the new speaker classifier, are jointly optimized.
\end{itemize}

\subsection{Semi-supervised Learning}

In this part, we investigate scenarios where the in-domain far-field data is unlabeled. In this case, the VoxCeleb data is consider as the out-of-domain labeled data and the FFSVC dataset is treated as the in-domain unlabeled data. The labeled data is used to pre-train a model for pseudo-labeling. For semi-supervised learning (SSL) operation in the transfer learning phase, the speaker embedding from unlabeled data is extracted using the pre-train model and the followed clustering algorithms to generate pseudo-labels for the unlabeled data. The following describes the algorithm generating pseudo-labels:

\begin{itemize}
    \item Step 1. Extract all speaker embeddings $\mathcal{Z} \in \mathbb{R}^{N\times d}$ from the FFSVC20 dataset using the pre-trained speaker model.
    \item Step 2. Run a clustering algorithm with the different number of clusters $K$ to obtain centroid matrix $\mathbf{C}\in \mathbb{R}^{d}$ for each $K$.
    \item Step 3. Calculate the within-class cosine similarity (WCCS) and observe the `elbow' of the WCCS curve to determine the number of clusters $K$ .
    \item Step 4. Create the pseudo labels for the FFSVC20 dataset.
    \item Step 5. Use the pseudo-labels data together with the labeled data into the speaker embedding model to fine-tune the model. 
    \item Step 6. Repeat Step 1 with the fine-tuned model from Step 5 as the pre-trained model.
\end{itemize}

\begin{figure}
  \centering
  \includegraphics[width=\linewidth]{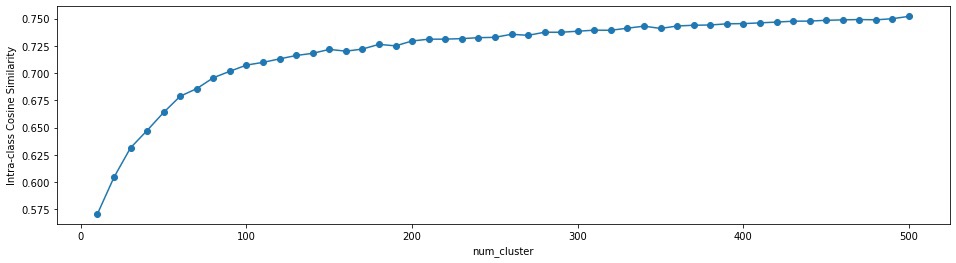}
  \caption{Within-Cluster Cosine Similarity versus the number of cluster K employed in FFSVC supplementary set}
  \label{fig:wccs}
\end{figure}

%\begin{table}[htp] 
%  \caption{Baseline system performance in FFSVC2022 evaluation set.}
%  \label{tab:baseline_result}
%  \centering
%  \begin{tabular}[c]{lcccccccc}
%    \toprule
%	\multirow{2}*{\textbf{Track}} &  \multicolumn{2}{c}{\textbf{Vox-O}} &  \multicolumn{2}{c}{\textbf{FFSVC Eval}} \\
%	\cmidrule(lr){2-3}  \cmidrule(lr){4-5} & \textbf{EER} & \textbf{mDCF} & \textbf{EER} & \textbf{mDCF} \\
%	\midrule
%	 Pre-train  & 2.07\% & 0.215 \\
%	  \quad +FT-mix (task1) & - & - & 7.021 & 0.681 \\
%	  \quad +SSL (task2) & - & - & 7.644 & 0.739 \\
%	\bottomrule
%     \end{tabular}
%\end{table}

\begin{table*}[htbp] 
  \caption{Final ranks for FFSVC2022 both tasks.}

  \label{tab:challenge_results}
  \centering
  \begin{tabular}[c]{ccccccc}
    \toprule
    \textbf{Track} & \textbf{Rank} &  \textbf{Team name} & \textbf{Organisation} & \textbf{EER} & \textbf{mDCF} \\
	
	\midrule
	\multirow{5}*{\bf{1}} & - & Baseline & FFSVC Organizer & 7.021 & 0.681 \\
					 & 4 & ZXIC \cite{zxic}    & ZTE Corporation & 4.409 & 0.511 \\
					 & 3 & Nan7U \cite{nan7u}   & - & 4.930 & 0.482 \\
					 & 2 & HiMia \cite{npu}   & Northwestern Polytechnical University, Huawei Cloud & 3.470 & 0.319 \\
					 & 1 & SPEAKIN \cite{speakin} & SpeakIn Technologies Co. Ltd. & 3.005 & 0.294 \\
					 
	\midrule
	\multirow{3}*{\bf{2}} & - & Baseline & FFSVC Organizer & 7.644 & 0.739 \\
					 & 2 & HiMia \cite{npu} & Northwestern Polytechnical University, Huawei Cloud & 5.342 & 0.545 \\
					 & 1 & SPEAKIN \cite{speakin} & SpeakIn Technologies Co. Ltd. & 6.692 & 0.537 \\
	\bottomrule
     \end{tabular}
\end{table*}

\textbf{Generating pseudo label by clustering}. We adopt the K-means algorithm as the clustering algorithm to generate the pseudo labels. The learning objective of K-means is set to minimize the within-cluster sum-of-squares criterion:
\begin{equation}
   \underset{\mathbf{C}}{min}\frac{1}{N}\sum_{i=1}^{N}\underset{k}{min}\Vert \mathbf{z}_{i}-\mathbf{C}_{k}\Vert^{2}
\end{equation}
where $\mathbf{z}_{i}\in \mathbb{R}^{d}$ is the $d$-dimensional speaker embedding of the $i^{th}$ sample. The cluster with the closest controid to $z_{i}$ in terms of the L2-norm distance is assigned as the pseudo-label for sample i. 

\textbf{Determine the number of clusters}. Inspired by the works of Cai \textit{et al.}\cite{dku_voxsrc20,dku_voxsrc21_ssl}, we determine the number of clusters by the `elbow' method. Given $\mathbf{z}_{k,a}$, the assigned $a^{th}$ embedding of the $k^{th}$ cluster. The total WCCS of $N$ elements is:  
\begin{equation}
	WCCS=\frac{\sum_{k=1}^{K}\sum_{a=1}^{A} cos(\mathbf{z}_{k,a},\mathbf{C}_{k}) }{N},
\end{equation}

Since the cosine similarity and euclidean similarity are connected linearly for normalized vectors, the WCCS linearly connects with learning objective of K-means. Fig. \ref{fig:wccs} shows the curve of WCCS results under different $K$s. The WCCS monotonically increases as number of clusters $K$ increases. WCCS tends to flatten with some $K$ onwards and forming an `elbow' of the curve. Such `elbow' indicates that the intra-cluster has little variation and  increasingly over-fitting. From Fig. \ref{fig:wccs}, the `elbow' is distributed between 100 and 120. The specific implementation of baseline system has been released\footnote{https://github.com/FFSVC/FFSVC2022\_Baseline\_System}.

\begin{table*}[bp] 
  \caption{Challenge results of text-mismatch and cross-time evaluation scenarios. TD and TI indicate that Text-dependent and Text-independent case. The Visit Gap 0/1/2 indicates that the visit time gap of enroll and test data is 0/1/2.}

  \label{tab:analysis}
  \centering
  \begin{tabular}[c]{cccccccccccccc}
    \toprule
    \multirow{2}*{\textbf{Track}} & \multirow{2}*{\textbf{User}} &  \multicolumn{2}{c}{\textbf{TD}} & \multicolumn{2}{c}{\textbf{TI}} &  \multicolumn{2}{c}{\textbf{Visit Gap 0}} & \multicolumn{2}{c}{\textbf{Visit Gap 1}} & \multicolumn{2}{c}{\textbf{Visit Gap 2}} & \multicolumn{2}{c}{\textbf{Final}} \\
     \cmidrule(lr){3-4} \cmidrule(lr){5-6} \cmidrule(lr){7-8} \cmidrule(lr){9-10} \cmidrule(lr){11-12} \cmidrule(lr){13-14} & & \textbf{EER} & \textbf{mDCF} & \textbf{EER} & \textbf{mDCF}& \textbf{EER} & \textbf{mDCF}& \textbf{EER} & \textbf{mDCF} & \textbf{EER} & \textbf{mDCF} & \textbf{EER} & \textbf{mDCF}\\
	
	\midrule
	\multirow{9}*{\textbf{Task 1}} &  Baseline & 4.427 & 0.464 &  7.11 & 0.69 & 6.066 & 0.608 & 7.630 & 0.734 & 7.938 & 0.749 & 7.021 & 0.681 \\
	  & A08 & 3.756 & 0.380 & 6.745 & 0.634 & 5.776 & 0.559 & 7.237 & 0.677 & 7.593 & 0.695 & 6.674 & 0.627 \\
	  & A07      & 3.968 & 0.393 & 6.940 & 0.637 & 4.317 & 0.454 & 8.464 & 0.780 & 7.935 & 0.713 & 6.846 & 0.626 \\
	  & A06     & 3.529 & 0.360 & 6.062 & 0.603 & 5.055 & 0.528 & 6.517 & 0.644 & 7.039 & 0.662 & 5.993 & 0.595 \\
      & A05        & 2.688 & 0.309 & 4.363 & 0.502 & 3.426 & 0.398 & 4.979 & 0.591 & 5.178 & 0.611 & 4.409 & 0.511 \\
	  & A04           & 2.250 & 0.193 & 4.961 & 0.489 & 4.027 & 0.411 & 5.446 & 0.534 & 5.974 & 0.554 & 4.930 & 0.482 \\
	  & A03        & 1.904 & 0.236 & 3.482 & 0.396 & 2.810 & 0.332 & 3.860 & 0.457 & 4.390 & 0.477 & 3.510 & 0.404 \\
	  & A02           & 1.832 & 0.169 & 3.510 & 0.321 & 2.765 & 0.265 & 3.776 & 0.347 & 4.445 & 0.401 & 3.470 & 0.319 \\
	  & A01      & 1.455 & 0.158 & 3.036 & 0.297 & 2.359 & 0.237 & 3.336 & 0.334 & 3.787 & 0.356 & 3.005 & 0.294 \\			
	  
	\midrule
	\multirow{3}*{\textbf{Task 2}} &  Baseline & 5.124 & 0.562 & 7.704 & 0.743 & 6.682 & 0.670 & 8.299 & 0.792 & 8.480 & 0.800 & 7.644 & 0.739 \\
	  & B02    & 3.081 & 0.334 & 5.405 & 0.552 & 4.478 & 0.469 & 5.794 & 0.600 & 6.387 & 0.623 & 5.342 & 0.545 \\
	  & B01  & 3.746 & 0.340 & 6.66 & 0.527 & 5.701 & 0.475 & 7.278 & 0.583 & 7.815 & 0.601 & 6.692 & 0.537 \\	 
					 
	\bottomrule
	
     \end{tabular}
\end{table*}

\subsection{Experimental setup}
	
The acoustic features are 80-dimensional log Mel-filterbank energies with a frame length of 25ms and hop size of 10ms. The extracted features are mean-normalized before feeding into the deep speaker network. The network structure contains three main components: a front-end pattern extractor, an encoder layer, and a back-end classifier. The ResNet34 \cite{resnet} model, and different residual blocks [32, 64, 128, 256], is employed as the front-end pattern extractor, the 256-dimensional fully connected layer following the global statistic pooling (GSP) based encoder layer is adopted as the speaker embedding layer. The ArcFace\cite{arcface} (s = 32, m = 0.2) is used as the classifier.

SGD optimizer is adopted to update model parameters, and we adopt the MultiStepLR as the learning rate (LR) decays strategy that decays the learning rate of each parameter group by 0.1 once the number of epoch reaches one of the milestones. The milestone epochs are 10, 20, and 30. In the pre-train stage, LR decreases from the initialized 0.1 to 0.0001 until its performance no longer decreases on the development set. In the fine-tuning stage, the LR is set to a fixed constant of 0.001. We adopt the MUSAN dataset\cite{musan} and RIR\_NOISE\cite{rir_noise} dataset to do data augmentation.

%\subsection{Result}
%
%The Table \ref{tab:baseline_result} presents the baseline system performance. For task1, the baseline system adopts the FT-mix training strategy and achieves 30\% related improvement than pre-train model. For task2, the model has only been iterated once with K-means method and without any additional operation. 

\begin{table*}[bp] 
  \caption{Challenge results of cross-distance evaluation scenarios. The -1.5M, 1M, 3M and 5M indicates the distance between speaker and recored devices. }

  \label{tab:analysis_distance}
  \centering
  \begin{tabular}[c]{cccccccccccccc}
    \toprule
    \multirow{2}*{\textbf{Track}} & \multirow{2}*{\textbf{User}} &  \multicolumn{2}{c}{\textbf{-1.5M}} & \multicolumn{2}{c}{\textbf{1M}} &  \multicolumn{2}{c}{\textbf{3M}} & \multicolumn{2}{c}{\textbf{5M}} & \multicolumn{2}{c}{\textbf{Final}} \\
     \cmidrule(lr){3-4} \cmidrule(lr){5-6} \cmidrule(lr){7-8} \cmidrule(lr){9-10} \cmidrule(lr){11-12} & & \textbf{EER} & \textbf{mDCF} & \textbf{EER} & \textbf{mDCF}& \textbf{EER} & \textbf{mDCF}& \textbf{EER} & \textbf{mDCF} & \textbf{EER} & \textbf{mDCF} \\
	
	\midrule
	\multirow{9}*{\textbf{Task 1}} &  Baseline & 7.712 & 0.74 & 6.203 & 0.623 & 7.280 & 0.692 & 6.513 & 0.635 & 7.021 & 0.681 \\
	  & A08 & 7.197 & 0.662 & 6.01 & 0.578 & 6.858 & 0.637 & 6.292 & 0.59 & 6.674 & 0.627 \\
	  & A07      & 7.180 & 0.624 & 6.010 & 0.578 & 6.858 & 0.637 & 6.292 & 0.590 & 6.846 & 0.626 \\
	  & A06     & 6.413 & 0.624 & 5.350 & 0.555 & 6.181 & 0.623 & 6.628 & 0.531 & 5.993 & 0.595 \\
      & A05        & 5.027 & 0.529 & 3.674 & 0.474 & 4.503 & 0.509 & 4.070 & 0.504 & 4.409 & 0.511 \\
	  & A04           & 5.228 & 0.512 & 4.515 & 0.478 & 5.043 & 0.486 & 4.746 & 0.432 & 4.930 & 0.482 \\
	  & A03        & 3.981 & 0.432 & 3.027 & 0.377 & 3.620 & 0.407 & 3.258 & 0.382 & 3.510 & 0.404 \\
	  & A02           & 3.937 & 0.354 & 2.820 & 0.240 & 3.642 & 0.334 & 3.043 & 0.291 & 3.470 & 0.319 \\
	  & A01      & 3.313 & 0.302 & 2.536 & 0.250 & 3.117 & 0.321 & 2.819 & 0.258 &3.005 & 0.294 \\			
	  
	\midrule
	\multirow{3}*{\textbf{Task 2}} &  Baseline  & 8.401 & 0.794 & 6.795 & 0.690 & 7.919 & 0.750 & 7.127 & 0.685 & 7.644 & 0.739 \\
	  & B02    & 5.875 & 0.572 & 4.757 & 0.502 & 5.507 & 0.551 & 4.920 & 0.516 & 5.342 & 0.545 \\
	  & B01  & 7.476 & 0.592 & 5.231 & 0.440 & 6.915 & 0.560 & 6.082 & 0.481 & 6.692 & 0.537 \\	 
					 
	\bottomrule
	
     \end{tabular}
\end{table*}

\section{Challenge results}
\subsection{Final Rank}
We received four system descriptions to elaborate and verify the correctness of their system, so the final rankings are shown in Table \ref{tab:challenge_results}. Teams, which do not submit system descriptions, are not considered in the final rankings. All system description has been published\footnote{https://ffsvc.github.io/publication}. The best performing SPEAKIN team for both tasks gives an mDCF of 0.294 and 0.537, followed by the HiMia team which gives an mDCF of 0.319 and 0.545 for two tasks. The other teams also exceeded the baseline system by more than 30\%. For Task1, all participants adopt the transfer learning method that uses VoxCeleb data in the pre-train stage and fine-tune with challenge data or a mix of challenge data and VoxCeleb data. Further, SPEAKIN team \cite{speakin} additional adopts the Large-Margin Fine-Tuning to improve model performance and Sub-Mean method to reduce the domain mismatch. HiMia team \cite{npu} adopt the two-stage transfer learning to maintain strong speaker discrimination ability of the pre-trained model in the in-domain data. In addition, they novelly adopt the model soup strategy to average the weights of multiple models in the score fusion stage. Nan7u team \cite{nan7u} adopts the ResNet and its variant, bidirectional multi-scale feature aggregation module and global-local information-based dynamic convolution neural, to build far-field speaker verification model. ZXIC team \cite{zxic} introduces a novel multi-reader domain adaption learning method to alleviate this mismatch impact.

\subsection{Analysis}

In this subsection, we will further analyze the task difficulty by the final submitted results. In task 1, eight teams exceeded the baseline system, and two teams had results that over form the task2 baseline system. Based on the difficulty of our trial design, we list three difficult scenarios in Table \ref{tab:analysis} and Table \ref{tab:analysis_distance}, the text-mismatch scenario, the cross-time scenario and cross-distance scenario. 

First, all trials are considered text-dependent (TD) and text-independent (TI) cases for the text-mismatch scenario. Although we do not directly distinguish the TD and TI tasks in this challenge, the TD results of all participants outperform than TI case by comparing TD and TI results in Table \ref{tab:analysis}. That also indicates that reducing text influence is important to improve system performance.

Second, the real speaker verification is time-varying. Usually, enroll and test audio are not recorded at the same time. In the cross-time scenario, all positive trials are split into three parts, visit gap zero/one/two times. For example, the visit gap 0 time indicates that the enrollment and test audio are recorded at the same visit, and the visit gap 1 time indicates that the recorded time has 1 visit time gap (7-15days). From the column of the terms of Visit GAP 0/1/2 in Table \ref{tab:analysis}, it's easily observed that with the visit gap increases, the system performance degrades. So how to alleviate the time-varying question is a new challenge, and we further focus on this question in the next challenge.

\begin{figure}
  \centering
  \includegraphics[width=\linewidth]{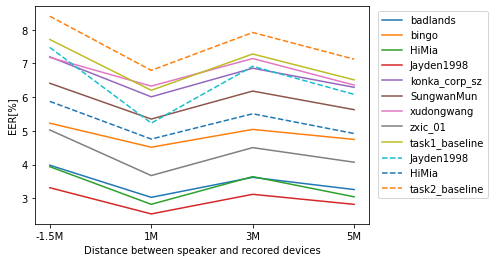}
  \caption{Challenge results of cross-distance evaluation scenarios. `$-$' and `$--$' indicate the task1 and task2 results, respectively.}
  \label{fig:result_distance}
\end{figure}

Finally, we focus on the cross-distance scenario. Table \ref{tab:analysis_distance} and Fig \ref{fig:result_distance} reports the results under different distance case. Since the speaker is directed sound sources, the results of -1.5M is the worst. Then, the result of 1M is best, and system performance degrades with increasing distance. In addition, since the 3M recorded device is closer to the noise source, performances of 3M are even worse than 5M.

\section{Conclusion}
This paper introduces the challenge task, evaluation protocol and baseline system. And we also further analyze the challenge results and participant method. This year, we focus on the single-channel far-field speaker verification, and two tasks are proposed: fully supervised speaker verification and semi-supervised speaker verification challenge.  

From the statistical analysis, far-field speaker verification is still a challenge and novel question. We designed a multi-scenario trial file consisting of text-mismatch, cross-time and cross distance for researchers. Researchers could reduce the impact of irrelevant speaker information and explore the domain adaptation strategy in this challenge.

\bibliographystyle{IEEEtran}

\bibliography{mybib}

\end{document}